\documentclass[aps,preprint,tightenlines,groupedaddress,nofootinbib]{revtex4}

\usepackage{graphicx,comment}

\usepackage{amssymb,latexsym}
\usepackage{amsmath,amsbsy}

\begin{document}

\unitlength=1mm

\def\a{{\alpha}}
\def\b{{\beta}}
\def\d{{\delta}}
\def\D{{\Delta}}
\def\e{{\epsilon}}
\def\g{{\gamma}}
\def\G{{\Gamma}}
\def\k{{\kappa}}
\def\l{{\lambda}}
\def\L{{\Lambda}}
\def\m{{\mu}}
\def\n{{\nu}}
\def\o{{\omega}}
\def\O{{\Omega}}
\def\S{{\Sigma}}
\def\s{{\sigma}}
\def\th{{\theta}}

\def\ol#1{{\overline{#1}}}

\def\Dslash{D\hskip-0.65em /}

\def\CPT{{$\chi$PT}}
\def\QCPT{{Q$\chi$PT}}
\def\PQCPT{{PQ$\chi$PT}}
\def\tr{\text{tr}}
\def\str{\text{str}}
\def\diag{\text{diag}}
\def\order{{\mathcal O}}

\def\cC{{\mathcal C}}
\def\cB{{\mathcal B}}
\def\cT{{\mathcal T}}
\def\cQ{{\mathcal Q}}
\def\cL{{\mathcal L}}
\def\cO{{\mathcal O}}
\def\cA{{\mathcal A}}

\def\eqref#1{{(\ref{#1})}}

\preprint{NT@UW-03-013}

\title{Charge Radii of the Meson and Baryon Octets
       in \\ Quenched and Partially Quenched Chiral Perturbation Theory}


\author{Daniel Arndt}
\email[]{arndt@phys.washington.edu}
\author{Brian C. Tiburzi}
\email[]{bctiburz@phys.washington.edu}
\affiliation{Department of Physics, Box 351560, 
             University of Washington, Seattle, WA 98195-1560, USA}

\date{\today}

\begin{abstract}
We calculate the electric charge radii 
of the $SU(3)$ pseudoscalar mesons and the $SU(3)$ octet baryons
in quenched and
partially quenched chiral perturbation theory.
We work in the isospin limit, 
up to next-to-leading order in the chiral expansion,
and to leading order in the heavy baryon expansion.
The results are necessary for the 
extrapolation of future lattice calculations of
meson and baryon charge radii.
We also derive expressions for the nucleon and pion charge radii 
in $SU(2)$ flavor
away from the isospin limit.
\end{abstract}



\maketitle

\section{Introduction}
The study of hadronic electromagnetic form factors at
low momentum transfer
provides important insight into the non-perturbative
structure of QCD.
A model-independent tool to study QCD at low energies is
chiral perturbation theory (\CPT), 
which is an effective field theory
with low-energy degrees of freedom, e.g., the meson octet
in $SU(3)$ flavor or the pion triplet in the case of $SU(2)$ flavor.
\CPT\ assumes that these mesons are the pseudo-Goldstone
bosons that appear from spontaneous breaking of chiral symmetry
from
$SU(3)_L\otimes SU(3)_R$ down to
$SU(3)_V$. 
Observables receive contributions from both
long-range and short range physics;
in \CPT\ the long-range contribution arises from
the  
(non-analytic) structure of 
pion loop contributions, while the short-range
contribution is encoded in a number of low-energy constants that
appear in the chiral Lagrangian and are unconstrained in
\CPT. 
These low-energy constants must
be determined from experiment or lattice simulations.

Notable progress toward measuring the 
proton and neutron form factors
has been made in recent years and high precision data 
are available 
(see \cite{Mergell:1996bf,Hammer:1996kx} for references).
Experimental study of the remaining octet baryons, however, 
is much harder.
The charge radius of the $\S^-$ has only recently been measured%
~\cite{Eschrich:1998tx}.
Although more experimental 
data for the other
baryon electromagnetic
observables
can be expected in the
future, 
progress will be slow as the experimental difficulties are significant.
Theory, however, may have a chance to catch up.
While quenched lattice calculations have already appeared%
~\cite{Tang:2003jh,Gockeler:2003ay,vanderHeide:2003ip,
Wilcox:1992cq,Draper:1990pi,Leinweber:1991dv},
with the advance of lattice gauge theory, 
we expect partially quenched calculations for many of these
observables
in the near future.
One problem that currently and foreseeably plagues 
these lattice calculations 
is that 
they cannot be performed with the physical masses of the light quarks.
Therefore, to make physical predictions,
it is necessary 
to extrapolate from the heavier quark masses
used on the lattice 
(currently on the order of the strange quark mass) down to
the physical light quark masses.
For lattice calculations that use the quenched approximation
of QCD (QQCD), 
where the fermion determinant that
arises from the path integral is set to one,  
quenched chiral perturbation theory (\QCPT)%
~\cite{Morel:1987xk,Sharpe:1992ft,Bernard:1992ep,
Bernard:1992mk,Golterman:1994mk,Sharpe:1996qp,Labrenz:1996jy}
has been developed to aid in
the extrapolation.  
The problem with the quenched approximation is
that the Goldstone boson singlet is no longer affected by
the $U(1)_A$ anomaly as in QCD.
In other words, the QQCD equivalent of the $\eta'$
that is heavy in QCD remains light and must be included in the
\QCPT\ Lagrangian.
This requires the addition of new operators and hence new
low-energy constants.
In general, the low-energy constants appearing in the \QCPT\
Lagrangian are unrelated to those in \CPT\ and
extrapolated quenched lattice data is unrelated to QCD.
In fact, several examples show that
the behavior of meson loops near the chiral limit is
frequently
misrepresented in \QCPT%
~\cite{Booth:1994rr,Savage:2001dy,Arndt:2002ed,Arndt:2003we}.
We find this is additionally true for the meson and 
baryon charge radii.

These problems can be remedied by using partially quenched 
lattice QCD (PQQCD).  
Unlike QQCD, where the masses of quarks not connected to
external sources are set to infinity,
these ``sea quark'' masses are kept finite in PQQCD.
The masses of the sea quarks can be varied independently
of the valence quark masses;
usually they are chosen to be heavier. 
Sea quarks are thereby kept as dynamical degrees of freedom
and the fermion determinant is no longer
equal to one. 
By efficaciously giving the sea quarks larger masses 
it is much less costly to calculate observables than
in ordinary QCD.
The low-energy effective theory of PQQCD is \PQCPT%
~\cite{Bernard:1994sv,Sharpe:1997by,Golterman:1998st,Sharpe:1999kj,
Sharpe:2000bn,Sharpe:2000bc,Sharpe:2001fh,Shoresh:2001ha}.
Since PQQCD retains a $U(1)_A$ anomaly,
the equivalent to the singlet field in QCD is heavy (on the order
of
the chiral symmetry breaking scale $\L_\chi$) and can be integrated out%
~\cite{Sharpe:2000bn,Sharpe:2001fh}.
Therefore, the low-energy constants appearing in \PQCPT\ are the same
as those appearing in \CPT.
By fitting \PQCPT\ to partially quenched
lattice data one can determine these constants  
and actually make physical predictions for QCD.
\PQCPT\ has been used recently to study 
heavy meson~\cite{Savage:2001jw} and
octet baryon observables%
~\cite{Chen:2001yi,Beane:2002vq,Savage:2002fm}.

While there are a number of lattice calculations for observables such
as the pion form factor%
~\cite{Martinelli:1988bh,Draper:1989bp,vanderHeide:2003ip}
or the octet baryon magnetic moments%
~\cite{Leinweber:1998ej,Hackett-Jones:2000qk}
that use the quenched approximation,
there are currently no partially quenched simulations.
However, 
given the recent progress that lattice gauge theory has made
in the one-hadron sector and the prospect of
simulations  
in the two-hadron sector%
~\cite{LATTICEproposal1,LATTICEproposal2,
Beane:2002np,Beane:2002nu,Arndt:2003vx},
we expect to see partially quenched calculations of the
electromagnetic
form factors in the near future.

The paper is organized as follows.
First, in Section~\ref{sec:PQCPT}, 
we review \PQCPT\ including the treatment
of the baryon octet and decuplet in the heavy baryon approximation.
In Section~\ref{sec:ff},
we calculate the charge radii of the 
meson and baryon octets in both \QCPT\ and \PQCPT\
up to next-to-leading (NLO) order in the chiral expansion.
We use the heavy baryon formalism of Jenkins and Manohar%
~\cite{Jenkins:1991jv,Jenkins:1991ne},
treat the decuplet baryons as dynamical degrees of freedom,
and keep contributions to lowest order in the heavy baryon mass, $M_B$.
These calculations are done in the 
isospin limit of $SU(3)$ flavor.  
For completeness we also provide the
\PQCPT\ 
result for the charge radii for the $SU(2)$ chiral
Lagrangian with non-degenerate quarks in the Appendix.
In Section~\ref{sec:conclusions} we conclude.

\section{\label{sec:PQCPT}\PQCPT}
In PQQCD the quark part of the Lagrangian is written as%
~\cite{Sharpe:2000bn,Sharpe:2001fh,Sharpe:2000bc,Sharpe:1999kj,
Golterman:1998st,Sharpe:1997by,Bernard:1994sv,Shoresh:2001ha}
\begin{eqnarray}\label{eqn:LPQQCD}
  {\cal L}
  &=&
  \sum_{a,b=u,d,s}\bar{q}_a(i\Dslash-m_q)_{ab} q_b
  + \sum_{\tilde{a},\tilde{b}=\tilde{u},\tilde{d},\tilde{s}}
      \bar{\tilde{q}}_{\tilde{a}}
      (i\Dslash-m_{\tilde{q}})_{\tilde{a}\tilde{b}} 
      \tilde{q}_{\tilde{b}}
  +
  \sum_{a,b=j,l,r}
  \bar{q}_{\text{sea},a} (i\Dslash-m_{\text{sea}})_{ab} q_{\text{sea},b}
               \nonumber \\
  &&=
  \sum_{j,k=u,d,s,\tilde{u},\tilde{d},\tilde{s},j,l,r}
  \bar{Q}_j(i\Dslash-m_Q)_{jk} Q_k
.\end{eqnarray}
Here, in addition to the fermionic light valence quarks $u$, $d$, and $s$ 
and their bosonic counterparts $\tilde{u}$, $\tilde{d}$, and $\tilde{s}$,
three light fermionic sea quarks $j$, $l$, and $r$ have been added.
These nine quarks are in the fundamental representation of
the graded group $SU(6|3)$%
~\cite{BahaBalantekin:1981kt,BahaBalantekin:1981qy,BahaBalantekin:1982bk}
and have been 
accommodated in the nine-component vector
\begin{equation}
  Q=(u,d,s,j,l,r,\tilde{u},\tilde{d},\tilde{s})
\end{equation}
that obeys the graded equal-time commutation relation
\begin{equation} \label{eqn:commutation}
  Q^\a_i({\bf x}){Q^\b_j}^\dagger({\bf y})
  -(-1)^{\eta_i \eta_j}{Q^\b_j}^\dagger({\bf y})Q^\a_i({\bf x})
  =
  \d^{\a\b}\d_{ij}\d^3({\bf x}-{\bf y})
,\end{equation}
where $\a$ and $\b$ are spin and $i$ and $j$ are flavor indices.
The graded equal-time commutation relations for two $Q$'s and two
$Q^\dagger$'s can be written analogously.
The grading factor 
\begin{equation}
   \eta_k
   = \left\{ 
       \begin{array}{cl}
         1 & \text{for } k=1,2,3,4,5,6 \\
         0 & \text{for } k=7,8,9
       \end{array}
     \right.
\end{equation}
takes into account the different statistics for
fermionic and bosonic quarks.
The quark mass matrix is given by 
\begin{equation}
  m_Q=\text{diag}(m_u,m_d,m_s,m_j,m_l,m_r,m_u,m_d,m_s)
\end{equation}
so that diagrams with closed ghost quark loops cancel 
those with valence quarks.
Effects of virtual quark loops are,
however, present due to the contribution of the finite-mass 
sea quarks. 
In the limit $m_j=m_u$, $m_l=m_d$, and $m_r=m_s$ QCD
is recovered.

It has been recently realized~\cite{Golterman:2001yv} that 
the light quark electric charge matrix $\cQ$ is not uniquely
defined in PQQCD.  
The only constraint one imposes is for
the charge matrix $\cQ$ to have vanishing
supertrace. Thus as in QCD, no new operators
involving the singlet component are subsequently introduced.
Following~\cite{Chen:2001yi} we use
\begin{equation}
  \cQ
  =
  \diag
  \left(
    \frac{2}{3},-\frac{1}{3},-\frac{1}{3},q_j,q_l,q_r,q_j,q_l,q_r
  \right)
\end{equation}
so that QCD is recovered in the limit 
$m_j\to m_u$, $m_l\to m_d$, and $m_r\to m_s$
independently of the $q$'s.

\subsection{Mesons}
For massless quarks,
the Lagrangian in Eq.~(\ref{eqn:LPQQCD}) exhibits a graded symmetry
$SU(6|3)_L \otimes SU(6|3)_R \otimes U(1)_V$ that is assumed 
to be spontaneously broken down to $SU(6|3)_V \otimes U(1)_V$. 
The low-energy effective theory of PQQCD that emerges by 
expanding about the physical vacuum state is \PQCPT.
The dynamics of the emerging 80~pseudo-Goldstone mesons 
can be described at lowest 
order in the chiral expansion by the $\order(E^2)$ Lagrangian%
\footnote{
Here, $E\sim p$, $m_\pi$ where $p$ is an external momentum.
}
\begin{equation}\label{eqn:Lchi}
  {\cal L} =
  \frac{f^2}{8}
    \str\left(D^\mu\Sigma^\dagger D_\mu\Sigma\right)
    + \l\,\str\left(m_Q\Sigma+m_Q^\dagger\Sigma^\dagger\right)
    + \a\partial^\mu\Phi_0\partial_\mu\Phi_0
    - \mu_0^2\Phi_0^2
\end{equation}
where
\begin{equation} \label{eqn:Sigma}
  \Sigma=\exp\left(\frac{2i\Phi}{f}\right)
  = \xi^2
,\end{equation}
\begin{equation}
  \Phi=
    \left(
      \begin{array}{cc}
        M & \chi^{\dagger} \\ 
        \chi & \tilde{M}
      \end{array}
    \right)
,\end{equation}
$f=132$~MeV,
and we have defined the gauge-covariant derivative
$D_\mu\S=\partial_\mu\S+ie\cA_\mu[\cQ,\S]$.
The str() denotes a supertrace over flavor indices.
The $M$, $\tilde{M}$, and $\chi$ are matrices
of pseudo-Goldstone bosons with quantum numbers of $q\ol{q}$ pairs,
pseudo-Goldstone bosons with quantum numbers of 
$\tilde{q}\ol{\tilde{q}}$ pairs, 
and pseudo-Goldstone fermions with quantum numbers of $\tilde{q}\ol{q}$ pairs,
respectively.
$\Phi$ is defined in the quark basis and normalized such that
$\Phi_{12}=\pi^+$ (see, for example, \cite{Chen:2001yi}).
Upon expanding the Lagrangian in \eqref{eqn:Lchi} one finds that
to lowest order
the mesons with quark content $Q\bar{Q'}$
are canonically normalized when
their masses are given by
\begin{equation}\label{eqn:mqq}
  m_{QQ'}^2=\frac{4\lambda}{f^2}(m_Q+m_{Q'})
.\end{equation}

The flavor singlet field given by $\Phi_0=\str(\Phi)/\sqrt{6}$
is, in contrast to the \QCPT\ case, rendered heavy by the $U(1)_A$
anomaly
and can therefore be integrated out in \CPT.
Analogously its mass $\mu_0$ can be taken to be 
on the order of the chiral symmetry breaking scale, 
$\mu_0\to\Lambda_\chi$.  
In this limit the 
flavor singlet propagator becomes independent of the
coupling $\a$ and 
deviates from a simple pole form~\cite{Sharpe:2000bn,Sharpe:2001fh}.

\subsection{Baryons}
Just as there are mesons in PQQCD with quark content
$\ol{Q}_iQ_j$ that contain 
valence, sea, and ghost quarks, there are baryons 
with quark compositions $Q_iQ_jQ_k$ that
contain all three types of quarks.
Restrictions on the baryon fields ${\mathcal B}_{ijk}$
come from the fact that these fields
must reproduce the familiar octet and decuplet
baryons when $i$, $j$, $k=1$-$3$~\cite{Labrenz:1996jy,Chen:2001yi}.
To this end, one decomposes the irreducible representations
of $SU(6|3)_V$ into 
irreducible representations of 
$SU(3)_{\text{val}} \otimes SU(3)_{\text{sea}} \otimes SU(3)_{\text{ghost}}
 \otimes U(1)$.
The method to construct the octet baryons is to use the
interpolating field
\begin{equation}
  \cB_{ijk}^\g
  \sim
  \left(Q_i^{\a,a}Q_j^{\b,b}Q_k^{\g,c}-Q_i^{\a,a}Q_j^{\g,c}Q_k^{\b,b}\right)
  \e_{abc}(C\g_5)_{\a\b}
,\end{equation}
which when restricted to $i$, $j$, $k=1$-$3$ 
has non-zero overlap with the octet baryons.
Using the commutation relations in Eq.~(\ref{eqn:commutation})
one sees that $\cB_{ijk}$ satisfies the symmetries
\begin{eqnarray}
  \cB_{ijk}&=&(-)^{1+\eta_j\eta_k}\cB_{ikj}, \nonumber \\
  0&=&\cB_{ijk}+(-)^{1+\eta_i\eta_j}\cB_{jik}
        +(-)^{1+\eta_i\eta_j+\eta_j\eta_k+\eta_k\eta_i}\cB_{kji}
                   \label{eqn:Bsymmetries}
.\end{eqnarray}  

The spin-1/2 baryon octet $B_{ijk}=\cB_{ijk}$,
where the
indices $i$, $j$, and $k$ are restricted to $1$-$3$,
is contained as a $(\bf 8,\bf 1,\bf1)$ of
$SU(3)_{\text{val}} \otimes SU(3)_{\text{sea}} \otimes SU(3)_{\text{ghost}}$
in the $\bf 240$ representation.
The octet baryons, written in the familiar two-index notation
\begin{equation}
  B=
    \left(
      \begin{array}{ccc}
        \frac{1}{\sqrt{6}}\L+\frac{1}{\sqrt{2}}\S^0 & \S^+ & p \\ 
        \S^- & \frac{1}{\sqrt{6}}\L-\frac{1}{\sqrt{2}}\S^0 & n \\
        \Xi^- & \Xi^0 & -\frac{2}{\sqrt{6}}\L
      \end{array}
    \right)
,\end{equation}
are embedded in $B_{ijk}$ as~\cite{Labrenz:1996jy}
\begin{equation}
  B_{ijk}
  =
  \frac{1}{\sqrt{6}}
  \left(
    \e_{ijl}B_{kl}+\e_{ikl}B_{jl}
  \right)
.\end{equation}

Besides the conventional octet baryons that contain valence quarks,
$qqq$,
there are also baryon fields with sea and ghost quarks
contained in the $\bf 240$, 
e.g., $qq_{\text{sea}}\tilde{q}$.  
Since we are only interested 
in calculating one-loop diagrams that have octet baryons
in the external states,
we will need only the $\cB_{ijk}$ 
of two valence and one sea quark or two valence and one ghost quark.
We use the construction in~\cite{Chen:2001yi}.

Similarly, 
the familiar spin-3/2 decuplet baryons are embedded
in the $\bf 165$.  
Here,
one uses the interpolating field
\begin{equation}
  \cT_{ijk}^{\a,\mu}
  \sim
  \left(
    Q_i^{\a,a}Q_j^{\b,b}Q_k^{\g,c}
    +Q_i^{\b,b}Q_j^{\g,c}Q_k^{\a,a}
    +Q_i^{\g,c}Q_j^{\a,a}Q_k^{\b,b}
  \right)
  \e_{abc}
  \left(C\g^\mu\right)_{\b\g}
\end{equation}
that describes the $\bf 165$ dimensional representation of $SU(6|3)_V$ 
and has non-zero overlap with the decuplet baryons when 
the indices are restricted
to $i$, $j$, $k=1$-$3$.
Due to the commutation relations in Eq.~\eqref{eqn:commutation},
$\cT_{ijk}$ satisfies the symmetries
\begin{equation} \label{eqn:Tsymmetries}
  \cT_{ijk}=(-)^{1+\eta_i\eta_j}\cT_{jik}=(-)^{1+\eta_j\eta_k}\cT_{ikj}
.\end{equation}

The decuplet baryons 
are then readily embedded in $\cT$ by construction:
$T_{ijk}=\cT_{ijk}$, where
the indices $i$, $j$, $k$ are restricted to $1$-$3$.
They transform as a $(\bf 10, \bf 1, \bf1)$ under
$SU(3)_{\text{val}} \otimes SU(3)_{\text{sea}} \otimes SU(3)_{\text{ghost}}$.
Because of Eq.~(\ref{eqn:Tsymmetries}), $T_{ijk}$ is
a totally symmetric tensor.  
Our normalization convention is such that $T_{111}=\D^{++}$.
For the spin-3/2 baryons consisting of two valence and one ghost quark
or two valence and one sea quark, we use the states constructed in%
~\cite{Chen:2001yi}.

At leading order in the heavy baryon expansion, the 
free Lagrangian for the $\cB_{ijk}$ and 
$\cT_{ijk}$ is given by~\cite{Labrenz:1996jy}
\begin{eqnarray} \label{eqn:L}
  {\mathcal L}
  &=&
  i\left(\ol\cB v\cdot{\mathcal D}\cB\right)
  +2\a_M\left(\ol\cB \cB{\mathcal M}_+\right)
  +2\b_M\left(\ol\cB {\mathcal M}_+\cB\right)
  +2\sigma_M\left(\ol\cB\cB\right)\str\left({\mathcal M}_+\right)
                              \nonumber \\
  &&-i\left(\ol\cT^\mu v\cdot{\mathcal D}\cT_\mu\right)
  +\D\left(\ol\cT^\mu\cT_\mu\right)
  +2\g_M\left(\ol\cT^\mu {\mathcal M}_+\cT_\mu\right)
  -2\ol\sigma_M\left(\ol\cT^\mu\cT_\mu\right)\str\left({\mathcal M}_+\right)
,\end{eqnarray}
where 
${\mathcal M}_+
  =\frac{1}{2}\left(\xi^\dagger m_Q \xi^\dagger+\xi m_Q \xi\right)$.
The brackets in (\ref{eqn:L}) are shorthands for field
bilinear invariants originally employed in~\cite{Labrenz:1996jy}.
To lowest order in the chiral expansion, Eq.~\eqref{eqn:L} gives the 
propagators
\begin{equation}
  \frac{i}{v\cdot k},\quad
  \frac{iP^{\mu\nu}}{v\cdot k-\D}
\end{equation}
for the spin-1/2 and spin-3/2 baryons, respectively.
Here, $v$ is the velocity and $k$ the residual momentum of the
heavy baryon which are related to the momentum $p$ by
$p=M_B v+k$.
$M_B$ denotes the (degenerate) mass of the octet baryons
and $\D$ the decuplet--baryon mass splitting.
The polarization tensor
\begin{equation}
  P^{\mu\nu}
  =
  \left(v^\mu v^\nu-g^{\mu\nu}\right)-\frac{4}{3}S^\mu S^\nu
\end{equation}
reflects the fact that the Rarita-Schwinger field 
$(\cT^\mu)_{ijk}$ contains both spin-1/2 and spin-3/2 pieces;
only the latter remain as propagating
degrees of freedom (see \cite{Jenkins:1991ne}, for example).

The Lagrangian 
describing the relevant interactions of the $\cB_{ijk}$ 
and $\cT_{ijk}$ 
with the pseudo-Goldstone mesons is
\begin{equation} \label{eqn:Linteract}
  {\cal L}
  =
  2\a\left(\ol{\cB}S^\mu \cB A_\mu\right)
  +
  2\b\left(\ol{\cB}S^\mu A_\mu \cB\right)
  +
  \sqrt{\frac{3}{2}}\cC
  \left[
    \left(\ol{\cT}^\nu A_\nu \cB\right)+\text{h.c.}
  \right]  
.\end{equation}
The axial-vector and vector meson fields $A^\mu$ and $V^\mu$
are defined by analogy to those in QCD:
\begin{equation}
  A^\mu=\frac{i}{2}
        \left(
          \xi\partial^\mu\xi^\dagger-\xi^\dagger\partial^\mu\xi
        \right)\quad\text{and}\quad
  V^\mu=\frac{1}{2}
        \left(
          \xi\partial^\mu\xi^\dagger+\xi^\dagger\partial^\mu\xi
        \right)
.\end{equation}
The latter appears in Eq.~\eqref{eqn:Linteract} in the
covariant derivatives of $\cB_{ijk}$ and $\cT_{ijk}$ 
that both have the form
\begin{equation}
  ({\mathcal D}^\mu \cB)_{ijk}
  =
  \partial^\mu \cB_{ijk}
  +(V^\mu)_{il}\cB_{ljk}
  +(-)^{\eta_i(\eta_j+\eta_m)}(V^\mu)_{jm}\cB_{imk}
  +(-)^{(\eta_i+\eta_j)(\eta_k+\eta_n)}(V^\mu)_{kn}\cB_{ijn}
.\end{equation}
The constants $\a$ and $\b$ are easily calculated in terms of the
constants $D$ and $F$ that are used for the $SU(3)_{\text{val}}$
analogs of these terms in QCD.  
Restricting the indices of $\cB_{ijk}$ to
$i,j,k=1,2,3$ one easily identifies
\begin{equation}
  \a=\frac{2}{3}D+2F\quad\text{and}\quad
  \b=-\frac{5}{3}D+F
.\end{equation}

\section{\label{sec:ff}Charge Radii}
In this Section we calculate the charge radii in
\PQCPT\ and \QCPT.
The basic conventions and notations for the
mesons and baryons in \PQCPT\ have been laid forth
in the last section; \QCPT\ has been extensively reviewed in 
the literature%
~\cite{Morel:1987xk,Sharpe:1992ft,Bernard:1992ep,
Bernard:1992mk,Golterman:1994mk,Sharpe:1996qp,Labrenz:1996jy}.%
\footnote{
In short, there are only valence and ghost quarks in the theory
and they have the same mass and charge pairwise so that diagrams with
disconnected quark loops are zero.
The singlet is light and has to be retained in the theory.
It is treated perturbatively and resulting observables generally
depend upon the parameters $\a$ and $\mu_0$.
}

\subsection{Octet Meson Charge Radii}
The electromagnetic form factor $G_{X}$ of an octet meson 
$\phi_X$
is required by
Lorentz invariance and gauge invariance to have the form
\begin{equation}\label{eqn:mesonff}
  \langle\phi_{X}(p')|J^\mu_{\text{em}}|\phi_{X}(p)\rangle
  = 
  G_{X}(q^2)(p+p')^\mu
\end{equation}
where
$q^\mu=(p'-p)^\mu$ 
and $p$ ($p'$) is the momentum of the incoming (outgoing) meson.
Conservation of electric charge protects it from
renormalization,
hence at
zero momentum transfer
$e G_X(0)=Q_X$, where $Q_X$ is the charge of $\phi_X$.
The charge radius $r_{X}$
is related to the slope of $G_{X}(q^2)$ at $q^2=0$,
namely
\begin{equation}
  <r_{X}^2>
  =
  6\frac{d}{dq^2}G_{X}(q^2)|_{q^2=0}
.\end{equation}

There are three terms in the $\order(E^4)$ Lagrangian
\begin{eqnarray} \label{eqn:L4PQQCD}
  {\cal L}
  &=&
  \a_4\frac{8\lambda}{f^2}
    \str(D_\mu\Sigma D^\mu\Sigma)
    \str(m_Q\Sigma+m_Q^\dagger\Sigma^\dagger)
  +
  \a_5\frac{8\lambda}{f^2}
    \str(D_\mu\Sigma D^\mu\Sigma(m_Q\Sigma+m_Q^\dagger\Sigma^\dagger))
                 \nonumber \\
  &&+
  i\a_9
    \str(L_{\mu\nu}D^\mu\Sigma D^\nu\Sigma^\dagger
                        +R_{\mu\nu}D^\mu\Sigma^\dagger D^\nu\Sigma)
  + \dots
\end{eqnarray}
that contribute to meson form factors at tree level.
Here $L_{\mu\nu}$, $R_{\mu\nu}$ are the field-strength tensors of 
the external sources, which for an
electromagnetic source are given by
\begin{eqnarray} \label{eqn:LR}
  L_{\mu\nu} = R_{\mu\nu}
  = e\cQ(\partial_\mu \cA_\nu-\partial_\nu \cA_\mu)+ie^2\cQ^2[\cA_\mu,\cA_\nu]
.\end{eqnarray}
Unlike \QCPT, where the low-energy constants are unique and have no known
connection to \CPT, in \PQCPT\ the parameters in
\eqref{eqn:L4PQQCD} are the dimensionless Gasser-Leutwyler
coefficients of \CPT~\cite{Gasser:1985gg}
which can be seen by looking at mesons that contain sea quarks only.

\begin{figure}[tb]
  \includegraphics[width=0.75\textwidth]{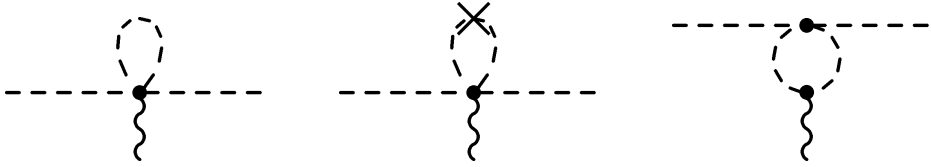}%
  \caption{
    Loop diagrams contributing to the octet meson charge radii
    in \PQCPT.
    Octet mesons are denoted by a dashed line,
    singlets (hairpins) by a crossed dashed line, 
    and the photon by a wiggly line.
    Only the third diagram has 
    $q^2$ dependence and therefore contributes to
    the charge radius.
  }
  \label{F:pions}
\end{figure}

\begin{figure}[tb]
  \includegraphics[width=0.50\textwidth]{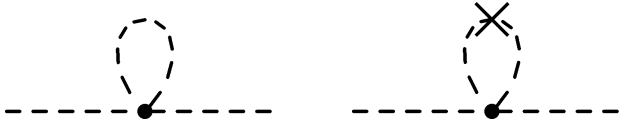}%
  \caption{
    Wavefunction renormalization diagrams  
    in \PQCPT.
    These diagrams, together with the third one in 
    Fig.~(\ref{F:pions}),
    ensure
    meson electric charge non-renormalization.
  }
  \label{F:pions-wf}
\end{figure}

To calculate the charge radii to lowest order
in the chiral expansion 
one has to include operators of $\cL$ in \eqref{eqn:Lchi}
to one-loop order 
[see Figs.~(\ref{F:pions}) and (\ref{F:pions-wf})]
and operators of \eqref{eqn:L4PQQCD}
to tree level. 
Using dimensional regularization, where
we have subtracted 
$\frac{1}{\e}+1-\gamma+\log 4\pi$,
we find in \PQCPT\ for the
$\pi^+$
\begin{eqnarray}\label{eqn:donaldduck}
  G^{PQ}_{\pi^+}(q^2)
  &=&
  1
  -
  \frac{1}{16\pi^2f^2}
  \left[2F_{uj}+F_{ur}\right]
  +\a_9\frac{4}{f^2}q^2
,\end{eqnarray}
which interestingly does not depend on the
charges of the sea and ghost quarks, $q_j$, $q_l$, $q_r$.
For the $K^+$ we find
\begin{eqnarray}
  G^{PQ}_{K^+}(q^2)
  &=&
  1
  +
  \frac{1}{16\pi^2f^2}
  \left[
    \left(\frac{1}{3}-q_{jl}\right)F_{uu}
    -
    \left(\frac{4}{3}-q_{jl}\right)F_{uj}
    -
    \left(\frac{2}{3}-q_r\right)F_{ur}    
    +
    \left(\frac{1}{3}+q_r\right)F_{ss}    
                  \right. \nonumber \\
     &&\phantom{dddddd} \left.
    -
    \left(\frac{2}{3}-q_{jl}+q_r\right)F_{us}    
    -
    \left(\frac{2}{3}+q_{jl}\right)F_{js}    
    -
    \left(\frac{1}{3}+q_r\right)F_{rs}    
  \right]
  +\a_9\frac{4}{f^2}q^2
\end{eqnarray}
and for the $K^0$ we find
\begin{eqnarray}\label{eqn:mickey}
  G^{PQ}_{K^0}(q^2)
  &=&
  \frac{1}{16\pi^2f^2}
  \left[
    \left(\frac{1}{3}-q_{jl}\right)F_{uu}
    +
    \left(\frac{2}{3}+q_{jl}\right)F_{uj}
    +
    \left(\frac{1}{3}+q_r\right)F_{ur}    
    +
    \left(\frac{1}{3}+q_r\right)F_{ss}    
                  \right. \nonumber \\
     &&\phantom{dddddd} \left.
    -
    \left(\frac{2}{3}-q_{jl}+q_r\right)F_{us}    
    -
    \left(\frac{2}{3}+q_{jl}\right)F_{js}    
    -
    \left(\frac{1}{3}+q_r\right)F_{rs}    
  \right]
.\end{eqnarray}
Here $q_{jl}=q_j+q_l$ and we have defined
\begin{equation}
  F_{QQ'}
  =
  \frac{q^2}{6}\log\frac{m_{QQ'}^2}{\mu^2}
  -
  m_{QQ'}^2{\mathcal F}\left(\frac{q^2}{m_{QQ'}^2}\right)
,\end{equation}
where the function ${\mathcal F}(a)$ is given by
\begin{equation}
   {\mathcal F}(a)
   = 
   \left(\frac{a}{6}-\frac{2}{3}\right)
   \sqrt{1-\frac{4}{a}}
   \log\frac{\sqrt{1-\frac{4}{a}+i\e}-1}{\sqrt{1-\frac{4}{a}+i\e}+1}
   +\frac{5a}{18}-\frac{4}{3}
.\end{equation}
The first derivative of $F_{QQ'}$ at $q^2=0$, needed
to calculate the charge radii, becomes
\begin{equation}
   6\frac{d}{dq^2}F_{QQ'}|_{q^2=0}
   =
   \log\frac{m_{QQ'}^2}{\mu^2}+1
.\end{equation}
Charge conjugation implies 
\begin{equation}
  G^{PQ}_{\pi^-}=-G^{PQ}_{\pi^+},\quad
  G^{PQ}_{K^-}=-G^{PQ}_{K^+},\quad\text{and}\quad
  G^{PQ}_{\ol{K}^0}=-G^{PQ}_{K^0}
,\end{equation}
which we have also verified at one-loop order.
The form factors of the flavor diagonal mesons are zero
by charge conjugation invariance.
In the limit $m_j\to\bar{m}$, $m_r\to m_s$ 
we recover the
QCD result~\cite{Gasser:1985gg,Gasser:1985ux}
as expected.

It is interesting to note,
that duplicating these calculations for 
\QCPT\ 
shows that there is no meson mass dependence at this order.
Specifically we find
\begin{equation}\label{eqn:GWB}
  G_{\pi^+}^Q(q^2)=-G_{\pi^-}^Q(q^2)
 =G_{K^+}^Q(q^2)=-G_{K^-}^Q(q^2)=1+\frac{4}{f^2}\a_9^Qq^2,
\end{equation}
and the form factors of the neutral mesons are zero.
Here we annotate the quenched constant $a_9^Q$ with a
``Q'' since its numerical value is different from the
one in Eq.~\eqref{eqn:L4PQQCD}.
Eq.~\eqref{eqn:GWB} reflects that 
flavor-singlet loops do not contribute to
the $q^2$-dependence at this order;
thus the virtual quark loops are completely removed
by their ghostly counterparts.
This can readily be seen by considering the quenched limit
of Eqs.~\eqref{eqn:donaldduck}--\eqref{eqn:mickey}.
The meson mass independence reveals once again
the pathologic nature of the quenched approximation
and seriously puts into question \CPT\ extrapolations to the
physical pion mass.

\subsection{Octet Baryon Charge Radii}
The electromagnetic form factors at or near zero momentum transfer
that enable the extraction of the baryon magnetic moments
and charge radii have been frequently investigated in QCD%
~\cite{Jenkins:1991jv,Jenkins:1993pi,
Meissner:1997hn,Bernard:1998gv,Kubis:1999xb,Kubis:2000aa,Kubis:2000zd,
Puglia:1999th,Puglia:2000jy,
Durand:1998ya}.
There are also recent quenched and partially quenched calculations
of the octet baryon magnetic moments in \QCPT\ and \PQCPT%
~\cite{Savage:2001dy,Chen:2001yi,Leinweber:2002qb}.
Here, we extend these calculations to the octet baryon charge radii.
We retain spin-3/2 baryons in intermediate states since 
formally $\D\sim m_\pi$.

Using the
heavy baryon formalism%
~\cite{Jenkins:1991jv,Jenkins:1991ne},
the baryon matrix element of the
electromagnetic current $J^\mu$ can be parametrized 
in terms of the Dirac and Pauli form factors $F_1$ and $F_2$,
respectively,
as
\begin{equation}
  \langle\ol B(p') \left|J^\mu\right|B(p)\rangle
  =
  \,\ol u(p')
  \left\{
    v^\mu F_1(q^2)+\frac{[S^\mu,S^\nu]}{M_B}q_\nu F_2(q^2)
  \right\}
  u(p)
\end{equation}
with $q=p'-p$.
The Sachs electric and magnetic form factors
defined as
\begin{eqnarray}
  G_E(q^2)&=&F_1(q^2)+\frac{q^2}{4M_B^2}F_2(q^2) \label{eqn:bob}\\
  G_M(q^2)&=&F_1(q^2)+F_2(q^2)
\end{eqnarray}
are particularly useful.
The baryon charge $Q$, 
electric charge radius $<r_E^2>$, and magnetic moment $\mu$
can be defined in terms of these form factors by
\begin{equation}
  Q=G_E(0),\quad
  <r_E^2>=\left.6\frac{d}{dq^2}G_E(q^2)\right|_{q^2=0},
                      \quad
  \text{and}\quad
  \mu=G_M(0)-Q
.\end{equation}
Here the baryon charge $Q$ is in units of $e$.

\subsubsection{\PQCPT}
\begin{figure}
  \includegraphics[width=0.75\textwidth]{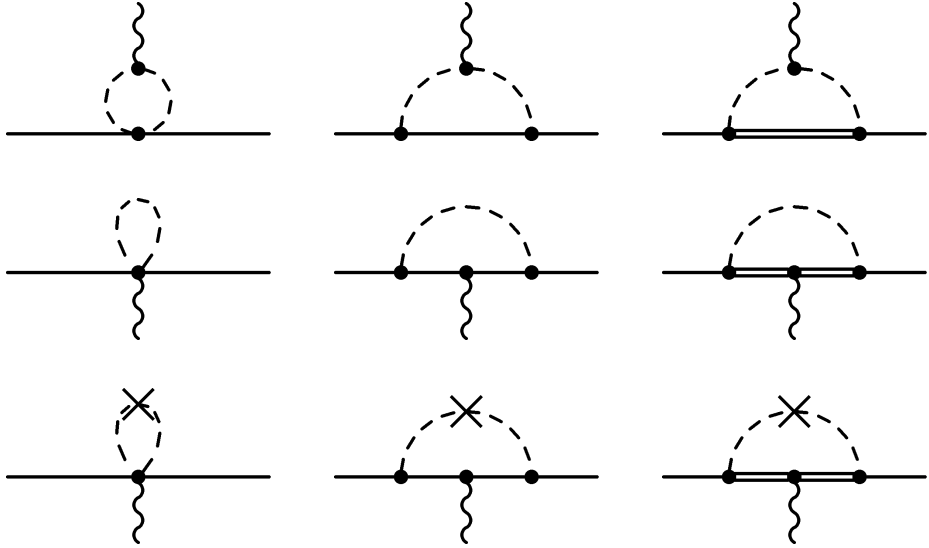}%
  \caption{\label{F:baryons}
     Loop diagrams contributing to the baryon magnetic moments
     and charge radii.
     A thin (thick) solid line denotes an octet (decuplet)
     baryon.
     The last two diagrams in the first row
     contribute to both magnetic moments and charge radii;
     the magnetic moment part of which has 
     already been calculated in \cite{Chen:2001yi}.
     The first diagram in row 1 contributes $q^2$ dependence
     only to $F_1$ 
     and therefore is relevant for the charge radii.
     The remaining diagrams have no $q^2$ dependence.
     These along with the wave function renormalization diagrams
     in Fig.~(\ref{F:baryons-wf}) maintain
     charge non-renormalization.}
\end{figure}

\begin{figure}
  \includegraphics[width=0.75\textwidth]{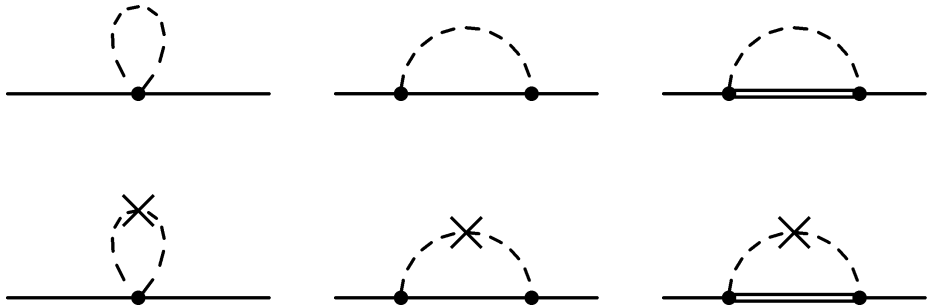}%
  \caption{\label{F:baryons-wf}
     Wave function renormalization diagrams
     needed to maintain
     baryon electric charge non-renormalization.}
\end{figure}

Let us first consider the calculation of the octet baryon charge radii
in \PQCPT.
Here, the leading tree-level correction to the magnetic moments 
come from the
dimension-5 operators%
\footnote{Here we use
$F_{\mu\nu}=\partial_\mu A_\nu-\partial_\nu A_\mu$.}
\begin{eqnarray}
  {\cal L}
  &=&
  \frac{ie}{2M_B}
  \left[
    \mu_\a\,\left(\ol{\cB}[S_\mu,S_\nu]\cB \cQ\right)
    +\mu_\b\,\left(\ol{\cB}[S_\mu,S_\nu]\cQ\cB\right)
  \right]
  F^{\mu\nu}
\end{eqnarray}
which
can be matched on the QCD Lagrangian
upon restricting the baryon field indices to 
$1$-$3$
\begin{eqnarray}\label{eqn:LDF}
  {\cal L}
  &=&
  \frac{ie}{2M_B}
  \left[
    \mu_D\,\tr(\ol{B}[S_\mu,S_\nu]\{\cQ,B\})
    +\mu_F\,\tr(\ol{B}[S_\mu,S_\nu][\cQ,B])
  \right]
  F^{\mu\nu}
\end{eqnarray}
where
\begin{equation}
  \mu_\a=\frac{2}{3}\mu_D+2\mu_F\quad\text{and }
  \mu_\b=-\frac{5}{3}\mu_D+\mu_F  
\end{equation}
at tree level.
The magnetic moments contribute the so-called Foldy term
to charge radii via $F_2(0)$ in Eq.~\eqref{eqn:bob}.
Likewise, further leading tree-level corrections to the charge radii
come from the dimension-6 operators
\begin{eqnarray}\label{eqn:Lc}
  {\cal L}
  &=&
  \frac{e}{\L_\chi^2}
  \left[
    c_\a\,\left(\ol{\cB}\cB\cQ\right)
    +c_\b\,\left(\ol{\cB}\cQ\cB\right)
  \right]
  v_\mu\partial_\nu F^{\mu\nu}
\end{eqnarray}
and the parameters $c_+$ and $c_-$,
defined by
\begin{equation}
  c_\a=\frac{2}{3}c_+ + 2c_-\quad\text{and }
  c_\b=-\frac{5}{3}c_+ + c_-  
,\end{equation}
are the same as those used in QCD.
Here, we take the chiral symmetry breaking scale 
$\L_\chi\sim 4\pi f$ for the purpose of power counting.
The NLO contributions arise from the one-loop diagrams shown in
Figs.~(\ref{F:baryons}) and (\ref{F:baryons-wf}).
To calculate the charge radii we need 
the form factors $F_1$ 
to first order in $q^2$ and 
$F_2(0)$
we find
\begin{eqnarray} \label{eqn:fred}
  <r_E^2>
  &=&
  -\frac{6}{\L_\chi^2}(Qc_-+\a_Dc_+)
  +
  \frac{3}{2M_B^2}(Q\mu_F+\a_D\mu_D)
                    \nonumber \\
  &&
  -
  \frac{1}{16\pi^2 f^2}
  \sum_{X}
  \left[ 
    A_X\log\frac{m_X^2}{\mu^2}
    -
    5\,\b_X\log\frac{m_X^2}{\mu^2}
    +
    10\,\b_X'{\mathcal G}(m_X,\D,\mu)
  \right]. \nonumber \\
\end{eqnarray}
Here, we have defined the function
${\mathcal G}(m,\D,\mu)$ 
by
\begin{equation}
  {\mathcal G}(m,\D,\mu)
  =
  \log\frac{m^2}{\mu^2}
  -\frac{\D}{\sqrt{\D^2-m^2}}
   \log\frac{\D-\sqrt{\D^2-m^2+i\e}}{\D+\sqrt{\D^2-m^2+i\e}}
.\end{equation}
Note that in Eq.~\eqref{eqn:fred} the only loop contributions
we keep are those non-analytic in $m_X$.
 
The parameters for the tree-level diagrams are listed in 
Table~\ref{T:treelevel}.
\begin{table}
\caption{\label{T:treelevel}Tree-level contributions in QCD, QQCD, and PQQCD}
\begin{tabular}{c | c c}\hline\hline
  & $Q$ & $\a_D$ \\ \hline
  $p$, $\S^+$ & 1 & $\frac{1}{3}$ \\
  $n$, $\Xi^0$ & 0 & $-\frac{2}{3}$\\
  $\S^0$ & 0 & $\frac{1}{3}$\\
  $\S^-$, $\Xi^-$ & $-1$ & $\frac{1}{3}$\\
  $\L$ & 0 & $-\frac{1}{3}$\\
  $\S^0\L$ & 0 & $\frac{1}{\sqrt{3}}$\\\hline\hline
\end{tabular}
\end{table}
The computed values for the $\b_X$, $\b_X'$, and $A_X$ coefficients
that appear in 
Eq.~(\ref{eqn:fred}) are listed for the 
octet baryons in 
Tables~\ref{T:p}---\ref{T:Lambda}.
\begin{table*}
\caption{\label{T:p}The coefficients $\b_X$, $\b_X'$, and $A_X$
         in $SU(3)$ flavor \PQCPT\ for the proton.}
\begin{ruledtabular}
\begin{tabular}{c | c c c}
  $X$ & $\b_X$ & $\b_X'$ & $A_X$\\ \hline
  $\pi$ & $-\frac{1}{9}\left(7D^2+6DF-9F^2\right)-\frac{1}{3}\left(5D^2-6DF+9F^2\right)q_{jl}$ & $\left(-\frac{2}{9}+\frac{1}{6}q_{jl}\right)\cC^2$ & $-1+3q_{jl}$\\
  $K$ & $-\frac{1}{9}\left(5D^2-6DF+9F^2\right)(1+3q_r)$ & $\left(\frac{1}{18}+\frac{1}{6}q_r\right)\cC^2$ & $1+3q_r$\\
  $uj$ & $-\frac{2}{9}\left(D+3F\right)^2+\frac{1}{3}\left(5D^2-6DF+9F^2\right)q_{jl}$ & $-\frac{1}{6}q_{jl}\cC^2$ & $2-3q_{jl}$\\
  $ur$ & $-\frac{1}{9}\left(D+3F\right)^2+\frac{1}{3}\left(5D^2-6DF+9F^2\right)q_r$ & $-\frac{1}{6}q_r\cC^2$ & $1-3q_r$
\end{tabular}
\end{ruledtabular}
\end{table*}
\begin{table*}
\caption{\label{T:n}The coefficients $\b_X$, $\b_X'$, and $A_X$
         in $SU(3)$ flavor \PQCPT\ for the neutron.}
\begin{ruledtabular}
\begin{tabular}{c | c c c}
  $X$ & $\b_X$ & $\b_X'$ & $A_X$\\ \hline
  $\pi$ & $\frac{1}{9}\left(17D^2-6DF+9F^2\right)-\frac{1}{3}\left(5D^2-6DF+9F^2\right)q_{jl}$ & $\left(\frac{1}{9}+\frac{1}{6}q_{jl}\right)\cC^2$ & $-1+3q_{jl}$\\
  $K$ & $-\frac{1}{9}\left(5D^2-6DF+9F^2\right)(1+3q_r)$ & $\left(\frac{1}{18}+\frac{1}{6}q_r\right)\cC^2$ & $1+3q_r$\\
  $uj$ & $-\frac{8}{9}\left(D^2-3DF\right)+\frac{1}{3}\left(5D^2-6DF+9F^2\right)q_{jl}$ & $\left(\frac{1}{9}-\frac{1}{6}q_{jl}\right)\cC^2$ & $-3q_{jl}$\\
  $ur$ & $-\frac{4}{9}\left(D^2-3DF\right)+\frac{1}{3}\left(5D^2-6DF+9F^2\right)q_r$ & $\left(\frac{1}{18}-\frac{1}{6}q_r\right)\cC^2$ & $-3q_r$
\end{tabular}
\end{ruledtabular}
\end{table*}
\begin{table*}
\caption{\label{T:Sigmaplus}The coefficients $\b_X$, $\b_X'$, and $A_X$
         in $SU(3)$ flavor \PQCPT\ for the $\S^+$.}
\begin{ruledtabular}
\begin{tabular}{c | c c c}
  $X$ & $\b_X$ & $\b_X'$ & $A_X$\\ \hline
  $\pi$ & $\frac{2}{9}\left(D^2+3F^2\right)(1-3q_{jl})$ & $\left(-\frac{1}{54}+\frac{1}{18}q_{jl}\right)\cC^2$ & $-\frac{2}{3}+2q_{jl}$\\
  $K$ & $-\frac{1}{9}\left(11D^2+6DF+3F^2\right)-(D-F)^2q_{jl}-\frac{2}{3}\left(D^2+3F^2\right)q_r$ & $\left(-\frac{5}{27}+\frac{1}{9}q_{jl}+\frac{1}{18}q_r\right)\cC^2$ & $\frac{1}{3}+q_{jl}+2q_r$\\
  $\eta_s$ & $-\frac{1}{3}(D-F)^2(1+3q_r)$ & $\left(\frac{1}{27}+\frac{1}{9}q_r\right)\cC^2$ & $\frac{1}{3}+q_r$\\
  $uj$ & $-\frac{2}{9}\left(D^2+3F^2\right)(4-3q_{jl})$ & $\left(\frac{2}{27}-\frac{1}{18}q_{jl}\right)\cC^2$ & $\frac{8}{3}-2q_{jl}$\\
  $ur$ & $-\frac{2}{9}\left(D^2+3F^2\right)(2-3q_r)$ & $\left(\frac{1}{27}-\frac{1}{18}q_r\right)\cC^2$ & $\frac{4}{3}-2q_r$\\
  $sj$ & $\frac{1}{3}(D-F)^2(2+3q_{jl})$ & $\left(-\frac{2}{27}-\frac{1}{9}q_{jl}\right)\cC^2$ & $-\frac{2}{3}-q_{jl}$\\
  $sr$ & $\frac{1}{3}(D-F)^2(1+3q_r)$ & $\left(-\frac{1}{27}-\frac{1}{9}q_r\right)\cC^2$ & $-\frac{1}{3}-q_r$
\end{tabular}
\end{ruledtabular}
\end{table*}
\begin{table*}
\caption{\label{T:Sigma0}The coefficients $\b_X$, $\b_X'$, and $A_X$
         in $SU(3)$ flavor \PQCPT\ for the $\S^0$.}
\begin{ruledtabular}
\begin{tabular}{c | c c c}
  $X$ & $\b_X$ & $\b_X'$ & $A_X$\\ \hline
  $\pi$ & $\frac{2}{9}\left(D^2+3F^2\right)(1-3q_{jl})$ & $\left(-\frac{1}{54}+\frac{1}{18}q_{jl}\right)\cC^2$ & $-\frac{2}{3}+2q_{jl}$\\
  $K$ & $-\frac{1}{9}\left(5D^2+6DF+3F^2\right)-(D-F)^2q_{jl}-\frac{2}{3}\left(D^2+3F^2\right)q_r$ & $\left(-\frac{11}{108}+\frac{2}{9}q_{jl}+\frac{1}{18}q_r\right)\cC^2$ & $\frac{1}{3}+q_{jl}+2q_r$\\
  $\eta_s$ & $-\frac{1}{3}(D-F)^2(1+3q_r)$ & $\left(\frac{1}{27}+\frac{1}{9}q_r\right)\cC^2$ & $\frac{1}{3}+q_r$\\
  $uj$ & $-\frac{2}{9}\left(D^2+3F^2\right)(1-3q_{jl})$ & $\left(\frac{1}{54}-\frac{1}{18}q_{jl}\right)\cC^2$ & $\frac{2}{3}-2q_{jl}$\\
  $ur$ & $-\frac{1}{9}\left(D^2+3F^2\right)(1-6q_r)$ & $\left(\frac{1}{108}-\frac{1}{18}q_r\right)\cC^2$ & $\frac{1}{3}-2q_r$\\
  $sj$ & $\frac{1}{3}(D-F)^2(2+3q_{jl})$ & $\left(-\frac{2}{27}-\frac{1}{9}q_{jl}\right)\cC^2$ & $-\frac{2}{3}-q_{jl}$\\
  $sr$ & $\frac{1}{3}(D-F)^2(1+3q_r)$ & $\left(-\frac{1}{27}-\frac{1}{9}q_r\right)\cC^2$ & $-\frac{1}{3}-q_r$
\end{tabular}
\end{ruledtabular}
\end{table*}
\begin{table*}
\caption{\label{T:Sigmaminus}The coefficients $\b_X$, $\b_X'$, and $A_X$
         in $SU(3)$ flavor \PQCPT\ for the $\S^-$.}
\begin{ruledtabular}
\begin{tabular}{c | c c c}
  $X$ & $\b_X$ & $\b_X'$ & $A_X$\\ \hline
  $\pi$ & $\frac{2}{9}\left(D^2+3F^2\right)(1-3q_{jl})$ & $\left(-\frac{1}{54}+\frac{1}{18}q_{jl}\right)\cC^2$ & $-\frac{2}{3}+2q_{jl}$\\
  $K$ & $\frac{1}{9}\left(D^2-6DF-3F^2\right)-(D-F)^2q_{jl}-\frac{2}{3}\left(D^2+3F^2\right)q_r$ & $\left(-\frac{1}{54}+\frac{1}{9}q_{jl}+\frac{1}{18}q_r\right)\cC^2$ & $\frac{1}{3}+q_{jl}+2q_r$\\
  $\eta_s$ & $-\frac{1}{3}(D-F)^2(1+3q_r)$ & $\left(\frac{1}{27}+\frac{1}{9}q_r\right)\cC^2$ & $\frac{1}{3}+q_r$\\
  $uj$ & $\frac{2}{9}\left(D^2+3F^2\right)(2+3q_{jl})$ & $\left(-\frac{1}{27}-\frac{1}{18}q_{jl}\right)\cC^2$ & $-\frac{4}{3}-2q_{jl}$\\
  $ur$ & $\frac{2}{9}\left(D^2+3F^2\right)(1+3q_r)$ & $\left(-\frac{1}{54}-\frac{1}{18}q_r\right)\cC^2$ & $-\frac{2}{3}-2q_r$\\
  $sj$ & $\frac{1}{3}(D-F)^2(2+3q_{jl})$ & $\left(-\frac{2}{27}-\frac{1}{9}q_{jl}\right)\cC^2$ & $-\frac{2}{3}-q_{jl}$\\
  $sr$ & $\frac{1}{3}(D-F)^2(1+3q_r)$ & $\left(-\frac{1}{27}-\frac{1}{9}q_r\right)\cC^2$ & $-\frac{1}{3}-q_r$
\end{tabular}
\end{ruledtabular}
\end{table*}
\begin{table*}
\caption{\label{T:Xi0}The coefficients $\b_X$, $\b_X'$, and $A_X$
         in $SU(3)$ flavor \PQCPT\ for the $\Xi^0$.}
\begin{ruledtabular}
\begin{tabular}{c | c c c}
  $X$ & $\b_X$ & $\b_X'$ & $A_X$\\ \hline
  $\pi$ & $\frac{1}{3}(D-F)^2\left(1-3q_{jl}\right)$ & $\left(-\frac{1}{27}+\frac{1}{9}q_{jl}\right)\cC^2$ & $-\frac{1}{3}+q_{jl}$\\
  $K$ & $\frac{1}{9}\left(11D^2+6DF+3F^2\right)-\frac{2}{3}\left(D^2+3F^2\right)q_{jl}-(D-F)^2q_r$ & $\left(\frac{5}{27}+\frac{1}{18}q_{jl}+\frac{1}{9}q_r\right)\cC^2$ & $-\frac{1}{3}+q_r+2q_{jl}$\\
  $\eta_s$ & $-\frac{2}{9}\left(D^2+3F^2\right)(1+3q_r)$ & $\left(\frac{1}{54}+\frac{1}{18}q_r\right)\cC^2$ & $\frac{2}{3}+2q_r$\\
  $uj$ & $-\frac{1}{3}(D-F)^2(4-3q_{jl})$ & $\left(\frac{4}{27}-\frac{1}{9}q_{jl}\right)\cC^2$ & $\frac{4}{3}-q_{jl}$\\
  $ur$ & $-\frac{1}{3}(D-F)^2(2-3q_r)$ & $\left(\frac{2}{27}-\frac{1}{9}q_r\right)\cC^2$ & $\frac{2}{3}-q_r$\\
  $sj$ & $\frac{2}{9}\left(D^2+3F^2\right)(2+3q_{jl})$ & $\left(-\frac{1}{27}-\frac{1}{18}q_{jl}\right)\cC^2$ & $-\frac{4}{3}-2q_{jl}$\\
  $sr$ & $\frac{2}{9}\left(D^2+3F^2\right)(1+3q_r)$ & $\left(-\frac{1}{54}-\frac{1}{18}q_r\right)\cC^2$ & $-\frac{2}{3}-2q_r$
\end{tabular}
\end{ruledtabular}
\end{table*}
\begin{table*}
\caption{\label{T:Ximinus}The coefficients $\b_X$, $\b_X'$, and $A_X$
         in $SU(3)$ flavor \PQCPT\ for the $\Xi^-$.}
\begin{ruledtabular}
\begin{tabular}{c | c c c}
  $X$ & $\b_X$ & $\b_X'$ & $A_X$\\ \hline
  $\pi$ & $\frac{1}{3}(D-F)^2\left(1+3q_{jl}\right)$ & $\left(-\frac{1}{27}+\frac{1}{9}q_{jl}\right)\cC^2$ & $-\frac{1}{3}+q_{jl}$\\
  $K$ & $-\frac{1}{9}\left(D^2-6DF-3F^2\right)-\frac{2}{3}\left(D^2+3F^2\right)q_{jl}-(D-F)^2q_r$ & $\left(\frac{1}{54}+\frac{1}{18}q_{jl}+\frac{1}{9}q_r\right)\cC^2$ & $-\frac{1}{3}+2q_{jl}+q_r$\\
  $\eta_s$ & $-\frac{2}{9}\left(D^2+3F^2\right)(1+3q_r)$ & $\left(\frac{1}{54}+\frac{1}{18}q_r\right)\cC^2$ & $\frac{2}{3}+2q_r$\\
  $uj$ & $\frac{1}{3}(D-F)^2(2+3q_{jl})$ & $\left(-\frac{2}{27}-\frac{1}{9}q_{jl}\right)\cC^2$ & $-\frac{2}{3}-q_{jl}$\\
  $ur$ & $\frac{1}{3}(D-F)^2(1+3q_r)$ & $\left(-\frac{1}{27}-\frac{1}{9}q_r\right)\cC^2$ & $-\frac{1}{3}-q_r$\\
  $sj$ & $\frac{2}{9}\left(D^2+3F^2\right)(2+3q_{jl})$ & $\left(-\frac{1}{27}-\frac{1}{18}q_{jl}\right)\cC^2$ & $-\frac{4}{3}-2q_{jl}$\\
  $sr$ & $\frac{2}{9}\left(D^2+3F^2\right)(1+3q_r)$ & $\left(-\frac{1}{54}-\frac{1}{18}q_r\right)\cC^2$ & $-\frac{2}{3}-2q_r$
\end{tabular}
\end{ruledtabular}
\end{table*}
\begin{table*}
\caption{\label{T:Lambda}The coefficients $\b_X$, $\b_X'$, and $A_X$
         in $SU(3)$ flavor \PQCPT\ for the $\L$.}
\begin{ruledtabular}
\begin{tabular}{c | c c c}
  $X$ & $\b_X$ & $\b_X'$ & $A_X$\\ \hline
  $\pi$ & $\frac{2}{27}\left(7D^2-12DF+9F^2\right)(1-3q_{jl})$ & $\left(-\frac{1}{18}+\frac{1}{6}q_{jl}\right)\cC^2$ & $-\frac{2}{3}+2q_{jl}$\\
  $K$ & $\frac{1}{27}\left(5D^2+30DF-9F^2\right)-\frac{1}{9}(D+3F)^2q_{jl}-\frac{2}{9}\left(7D^2-12DF+9F^2\right)q_r$ & $\left(\frac{5}{36}+\frac{1}{6}q_r\right)\cC^2$ & $\frac{1}{3}+q_{jl}+2q_r$\\
  $\eta_s$ & $-\frac{1}{27}(D+3F)^2(1+3q_r)$ & $0$ & $\frac{1}{3}+q_r$\\
  $uj$ & $-\frac{2}{27}\left(7D^2-12DF+9F^2\right)(1-3q_{jl})$ & $\left(\frac{1}{18}-\frac{1}{6}q_{jl}\right)\cC^2$ & $\frac{2}{3}-2q_{jl}$\\
  $ur$ & $-\frac{1}{27}\left(7D^2-12DF+9F^2\right)(1-6q_r)$ & $\left(\frac{1}{36}-\frac{1}{6}q_r\right)\cC^2$ & $\frac{1}{3}-2q_r$\\
  $sj$ & $\frac{1}{27}(D+3F)^2(2+3q_{jl})$ & $0$ & $-\frac{2}{3}-q_{jl}$\\
  $sr$ & $\frac{1}{27}(D+3F)^2(1+3q_r)$ & $0$ & $-\frac{1}{3}-q_r$
\end{tabular}
\end{ruledtabular}
\end{table*}
The corresponding values for the $\L\S^0$ transition are given in
Table~\ref{T:LambdaSigma0}.
\begin{table*}
\caption{\label{T:LambdaSigma0}The coefficients $\b_X$, $\b_X'$, and $A_X$
         in $SU(3)$ flavor \PQCPT\ for the $\L\S^0$ transition.}
\begin{ruledtabular}
\begin{tabular}{c | c c c}
  $X$ & $\b_X$ & $\b_X'$ & $A_X$\\ \hline
  $\pi$ & $-\frac{4}{3\sqrt{3}}D^2$ & $-\frac{1}{6\sqrt{3}}\cC^2$ & $0$\\
  $K$ & $-\frac{2}{3\sqrt{3}}D^2$ & $-\frac{1}{12\sqrt{3}}\cC^2$ & $0$\\
  $uj$ & $\frac{4}{3\sqrt{3}}D(D-3F)$ & $-\frac{1}{6\sqrt{3}}\cC^2$ & $0$\\
  $ur$ & $\frac{2}{3\sqrt{3}}D(D-3F)$ & $-\frac{1}{12\sqrt{3}}\cC^2$ & $0$
\end{tabular}
\end{ruledtabular}
\end{table*}
In each table we have listed the values corresponding to the 
loop meson that has mass $m_X$.  
If a particular meson is not listed then
the values for $\b_X$, $\b_X'$, and $A_X$ are zero.%
\footnote{
We have defined the coefficients $\b_X$ and
${\b_X}'$ to correspond to those defined in%
~\cite{Chen:2001yi} where 
$\mu=Q\,\mu_F+\a_D\,\mu_D
+\frac{M_B}{4\pi f^2}\sum_X\left[\b_X m_X
          +\b'_X{\mathcal F}(m_X,\D,\mu)\right]$
and the function ${\mathcal F}(m_X,\D,\mu)$ is
given in~\cite{Chen:2001yi}.
}

\subsubsection{\QCPT}
The calculation of the charge radii can be easily 
executed for \QCPT.
The operators in Eqs.~\eqref{eqn:LDF} and \eqref{eqn:Lc}
contribute, however, their low-energy coefficients
cannot be matched onto QCD.  Therefore we annotate them with a ``Q''.    
Additional terms involving hairpins~\cite{Labrenz:1996jy,Savage:2001dy}
do not contribute as 
their contribution to the charge radii is of the form
$(\mu_0^2/\L_\chi^4)\log m_q$
and therefore of higher order
in the chiral expansion.
We find
\begin{eqnarray} \label{eqn:fredQ}
  <r_E^2>
  &=&
  -\frac{6}{\L_\chi^2}(Qc^Q_-+\a_Dc^Q_+)
  +
  \frac{3}{2M_B^2}(Q\mu^Q_F+\a_D\mu^Q_D)
                    \nonumber \\
  &&
  +
  \frac{1}{16\pi^2 f^2}
  \sum_{X}
  \left[
    5\,\b^Q_X\log\frac{m_X^2}{\mu^2}
    -
    10\,{\b^Q_X}'{\mathcal G}(m_X,\D,\mu)
  \right]
.\end{eqnarray}
As with the meson case, the diagram where the photon couples 
to the closed meson loop does not contribute to baryon charge radii in the 
quenched case, {\it cf}., $A_X$ is zero.
The remaining coefficients appearing in Eq.~\eqref{eqn:fredQ} are 
listed in Table~\ref{T:QQCD}
\begin{table}
\caption{\label{T:QQCD}The coefficients $\b_X^Q$ and ${\b_X^Q}'$
         in $SU(3)$ flavor \QCPT\ for the octet baryons.}  
\begin{tabular}{l |  c  c  | c  c  }
	\hline
	\hline
	     & \multicolumn{2}{c |}{$\beta^Q_X$} &    \multicolumn{2}{c}{${\beta^Q_X}'$} \\
	 &   $\pi$     &   $K$  &  $\pi$  &   $K$    \\
        \hline
	$p$        	& $-\frac{4}{3}D_Q^2$ 	& $0$ 			& $-\frac{1}{6}\mathcal{C}_Q^2$ 	& $0$ \\
	$n$     	& $\frac{4}{3}D_Q^2$ 	& $0$ 			& $\frac{1}{6}\mathcal{C}_Q^2$ 	& $0$ \\
	$\Sigma^+$     	& $0$ 			& $-\frac{4}{3} D_Q^2$ 	& $0$ 			& $-\frac{1}{6}\mathcal{C}_Q^2$ \\
	$\Sigma^0$     	& $0$ 			& $-\frac{2}{3} D_Q^2$ 	&$0$ 			& $-\frac{1}{12}\mathcal{C}_Q^2$ \\
	$\Sigma^-$     	& $0$ 			& $0$ 			& $0$ 			& $0$ \\
	$\Lambda$     	& $0$ 			& $\frac{2}{3} D_Q^2$ 	& $0$ 			& $\frac{1}{12}\mathcal{C}_Q^2$ \\
	$\Xi^-$     	& $0$ 			& $0$			& $0$ 			& $0$ \\
	$\Xi^0$     	& $0$ 			& $\frac{4}{3} D_Q^2$ 	& $0$ 			& $\frac{1}{6}\mathcal{C}_Q^2$ \\
$\Sigma \Lambda$	& $-\frac{4}{3 \sqrt{3}} D_Q^2$ & $-\frac{2}{3 \sqrt{3}} D_Q^2$ & $-\frac{1}{6\sqrt{3}}\mathcal{C}_Q^2$ & $-\frac{1}{12\sqrt{3}}\mathcal{C}_Q^2$ \\
	\hline
	\hline
\end{tabular}
\end{table} 
and stem from meson loops formed solely
from valence quarks.

\section{\label{sec:conclusions}Conclusions}
We have calculated the charge radii for the octet mesons and baryons 
in the isospin limit of \PQCPT\
and also derive the result for
the nucleon doublet and pion triplet 
away from the isospin limit for the $SU(2)$
chiral Lagrangian.
For the octet mesons and baryons 
we have also calculated the \QCPT\ results.

Knowledge of the low-energy behavior of QQCD and PQQCD is crucial
to properly extrapolate lattice calculations from 
the quark masses used on the lattice to those in nature.
For the quenched approximation,
where the quark determinant is not evaluated,
one uses \QCPT\ to do this extrapolation.  
QQCD, however, has no known
connection to QCD and is
of merely academic interest.
Observables calculated in QQCD are often
found to be more divergent in the chiral limit than those in QCD.
This behavior is due to new operators included
in the QQCD Lagrangian, which are non-existent in QCD.
For the octet meson and baryon
charge radii we find that such operators enter at NNLO.
Hence, formally 
our NLO result is not more divergent than its QCD counterpart.
This, however, does not mean that our result is free of quenching
artifacts.
While the expansions about the chiral limit for QCD and QQCD
are formally similar,
$<r^2>\sim\a+\b\log m_Q+\dots$,
the QQCD result is anything but free of quenched oddities:
for certain baryons, $\S^-$ and $\Xi^-$ in particular,
diagrams that have bosonic or fermionic mesons running in loops 
completely cancel
so that $\b=0$.  In other words,
$<r^2>\sim\a+\dots$ and the result is actually independent of $m_Q$!
The same behavior is found for the charge radii of all mesons
in QQCD as the meson loop contributions entirely cancel.

PQQCD, on the other hand, is free of such eccentric behavior.
The formal behavior of the charge radius in the chiral limit
has the same form
as in QCD.
Moreover, 
there is 
a well-defined connection to QCD and one can
reliably extrapolate lattice results down to the 
quark masses of reality. 
The low-energy constants appearing in
PQQCD are the same as those in QCD and
by fitting them in \PQCPT\ one can make predictions for QCD.
Our \PQCPT\ result will 
enable the proper extrapolation of PQQCD lattice simulations
of the charge radii 
and we hope it encourages such simulations in the future.

\begin{acknowledgments}
We would like to thank Martin Savage 
for very helpful discussions and for useful comments on the manuscript.
This work is supported in part by the U.S.\ Department of Energy
under Grant No.\ DE-FG03-97ER4014.
\end{acknowledgments}

\appendix*
\section{\label{sec:SU2}Charge Radii in $SU(2)$ flavor
         with non-degenerate quarks}
In this section, we consider the case of $SU(2)$ 
flavor and calculate 
charge radii for the pions and
nucleons.
We keep the up and down quark masses non-degenerate and similarly 
for the sea-quarks. Thus the quark mass matrix reads
$m_Q^{SU(2)} = \diag(m_u, m_d, m_j, m_l, m_u, m_d)$.
Defining ghost and sea quark charges is constrained only by the 
restriction that QCD be recovered
in the limit of appropriately degenerate quark masses. 
Thus the most general form of the charge matrix is
\begin{equation}
  \cQ^{SU(2)} 
  = \diag\left(\frac{2}{3},-\frac{1}{3},q_j,q_l,q_j,q_l \right) 
.\end{equation}
The symmetry breaking pattern is assumed to be 
$SU(4|2)_L \otimes SU(4|2)_R \otimes U(1)_V
 \longrightarrow
 SU(4|2)_V \otimes U(1)_V$.

For the $\pi^+$, $\pi^-$, and $\pi^0$ we find
\begin{eqnarray}
  G^{PQ}_{\pi^+}(q^2)
  &=&
  1
  +
  \frac{1}{16\pi^2f^2}
  \left[
    \left(\frac{1}{3}+q_l\right)F_{dd}
    +
    \left(\frac{2}{3}-q_j\right)F_{uu}
    -
    \left(1-q_j+q_l\right)F_{ud}    
    -
    \left(\frac{1}{3}+q_j\right)F_{jd}    
                  \right. \nonumber \\
     &&\phantom{dddddd} \left.
    -
    \left(\frac{1}{3}+q_l\right)F_{ld}    
    -
    \left(\frac{2}{3}-q_j\right)F_{ju}    
    -
    \left(\frac{2}{3}-q_l\right)F_{lu}    
  \right]
  +\a_9\frac{4}{f^2}q^2
,\end{eqnarray}
$G^{PQ}_{\pi^-}=-G^{PQ}_{\pi^+}$, and $G^{PQ}_{\pi^0}=0$,
respectively.

The baryon field assignments are analogous to the case of
$SU(3)$ flavor.
The nucleons are embedded as
\begin{equation} \label{eqn:SU2nucleons}
 \cB_{ijk}=\frac{1}{\sqrt{6}}\left(\e_{ij} N_k + \e_{ik} N_j\right)
,\end{equation}
where the indices $i,j$ and $k$ are restricted to $1$ or $2$ and 
the $SU(2)$ nucleon doublet is defined as
\begin{equation}
  N = \left(\begin{matrix} p \\ n \end{matrix} \right) 
\end{equation}
The decuplet field $\cT_{ijk}$, 
which is totally symmetric, 
is normalized to contain the $\Delta$ resonances 
$T_{ijk}=\cT_{ijk}$ with $i$, $j$, $k$ restricted to 1 or 2.
The spin-3/2 baryon quartet is then contained as
\begin{equation}
  \cT_{111} = \D^{++}, \quad 
  \cT_{112} = \frac{1}{\sqrt{3}} \Delta^+, \quad 
  \cT_{122} = \frac{1}{\sqrt{3}} \Delta^0, \quad \text{and }
  \cT_{222} = \Delta^-
.\end{equation} 
The construction of the octet and decuplet baryons containing 
one sea or one ghost quark is analogous to the $SU(3)$ flavor
case~\cite{Beane:2002vq}
and we will not repeat it here.

The free Lagrangian for $\cB$ and $\cT$ is the one in 
Eq.~(\ref{eqn:Linteract})
(with the parameters having different numerical values than in
the $SU(3)$ case).  
The connection to QCD is 
detailed in~\cite{Beane:2002vq}.
Similarly, the Lagrangian describing the interaction of the 
$\cB$ and $\cT$ with the pseudo-Goldstone bosons is 
the one in 
Eq.~(\ref{eqn:Linteract}).  Matching it to the familiar one in QCD
(by restricting the $\cB_{ijk}$ and $\cT_{ijk}$ to the
$qqq$ sector)
\begin{equation}
  \cL = 2 g_A \ol{N} S^\mu A_\mu N 
      + g_1 \ol{N} S^\mu N \tr ( A_\mu) 
      + g_{\D N} \left( \ol{T}^{kji}_{\nu} A_{il}^{\nu} N_j \e_{kl} 
                      + \text{h.c.}  
                 \right)
\end{equation}
one finds at tree-level
\begin{equation}
  \a = \frac{4}{3} g_A + \frac{1}{3} g_1, \quad 
  \b = \frac{2}{3} g_1 - \frac{1}{3} g_A, \quad \text{and}\quad
  {\mathcal C} = - g_{\D N}
.\end{equation}

The contribution at leading order to the charge radii from the 
Pauli form factor $F_2(q^2)$, involves only the magnetic moments
which arise from the PQQCD Lagrangian~\cite{Beane:2002vq}
\begin{eqnarray}
  \cL &=& \frac{ie}{2 M_N}
        \left[ 
          \mu_\a\left(\ol\cB [S_\mu,S_\nu]\cB \cQ^{SU(2)}\right) 
         +\mu_\b\left(\ol\cB [S_\mu,S_\nu]\cQ^{SU(2)}\cB\right)
             \right.\nonumber \\
      &&\left.\phantom{ascuasd}
         +\mu_\g\,\str(\cQ^{SU(2)})\left(\ol\cB [S_\mu,S_\nu]\cB\right)
        \right]
        F^{\mu\nu}
.\end{eqnarray} 
Note that in the case of
$SU(2)$ flavor the charge matrix $\cQ$ is not
supertraceless and hence there appears a third operator.
In QCD, the corresponding Lagrange density is conventionally written 
in terms of isoscalar and isovector couplings
\begin{equation}
  \cL = \frac{ie}{2 M_N} 
        \left( 
          \mu_0 \ol N [S_\mu,S_\nu] N 
          + \mu_1 \ol N [S^\mu,S^\nu] \tau^3 N 
        \right)
        F^{\mu\nu}  
\end{equation}
and one finds that the 
QCD and PQQCD coefficients are related by~\cite{Beane:2002vq}
\begin{equation}
  \mu_0 = \frac{1}{6}\left(\mu_\a +\mu_\b +2\mu_\g\right), \quad \text{and }
  \mu_1 = \frac{1}{6}\left(2\mu_\a - \mu_\b\right)
.\end{equation}
Likewise, the leading tree-level corrections to the charge-radii come
from the Lagrangian
\begin{equation}
  \cL
  =
  \frac{e}{\L_\chi^2}
  \left[
    c_\a\,(\ol \cB \cB \cQ^{SU(2)})
    +c_\b\,(\ol \cB \cQ^{SU(2)}\cB)
    +c_\g\,\str(\cQ^{SU(2)})(\ol \cB \cB)
  \right]
  v_\mu\partial_\nu F^{\mu\nu}
\end{equation}
that matches onto the QCD Lagrangian
\begin{equation}
  \cL
  =
  \frac{e}{\L_\chi^2}
  \left(
    c_0\,\ol N N
    +c_1\,\ol N\tau^3 N
  \right)
  v_\mu\partial_\nu F^{\mu\nu}
\end{equation}
with
\begin{equation}
  c_0 = \frac{1}{6}\left(c_\a +c_\b +2c_\g\right), \quad \text{and }
  c_1 = \frac{1}{6}\left(2c_\a - c_\b\right)
.\end{equation}

Evaluating the charge radii at NLO order in the 
chiral expansion yields
\begin{equation}\label{eqn:fred2}
  <r_E^2>
  =
  -\frac{6c}{\L_\chi^2}
  +
  \frac{3\a}{2M_N^2}
  -
  \frac{1}{16\pi^2 f^2}
  \sum_{X}
  \left[ 
    A_X\log\frac{m_X^2}{\mu^2}
    -
    5\,\b_X\log\frac{m_X^2}{\mu^2}
    +
    10\,\b_X'{\mathcal G}(m_X,\D,\mu)
  \right]. 
\end{equation}
The coefficients $c$ are given by $c_p =c_0+c_1$ and $c_n =c_0-c_1$ 
while $\a_p = \mu_0 + \mu_1$ and $\a_n = \mu_0 - \mu_1$. 
The remaining coefficients are listed in Table~\ref{T:SU2p} for the proton 
and Table~\ref{T:SU2n} for the neutron.

\begin{table*}
\caption{\label{T:SU2p}The coefficients $\b_X$, $\b_X'$, and $A_X$
         in $SU(2)$ flavor \PQCPT\ for the proton.}
\begin{ruledtabular}
\begin{tabular}{c | c  c  c }
	$X$     &   $\beta_X$   &   $\beta'_X$   &  $A_X$  \\
	\hline
	$uu$   &   $\frac{2}{9}(4 g_A^2 + 2 g_1 g_A + g_1^2) (2 - 3 q_j)$  
						      &     $ (-\frac{1}{27} + \frac{1}{18} q_j)g_{\Delta N}^2$     &    $-\frac{4}{3} + 2 q_j$ \\	
	
	$ud$   &   $-\frac{4}{9} g_A^2 ( 5 + 6 q_l) + \frac{4}{9} g_1 g_A (2 - 3 q_l) + \frac{1}{9} g_1^2(1 - 9 q_j - 6 q_l)$           
						      &     $ (- \frac{2}{9} + \frac{1}{9} q_j + \frac{1}{18}q_l)g_{\Delta N}^2$  &    $q_j + 2 q_l$      \\
	
	$dd$   &   $-\frac{1}{3} g_1^2 ( 1 + 3 q_l)$   &     $( \frac{1}{27} + \frac{1}{9} q_l)g_{\Delta N}^2$     &    $\frac{1}{3} + q_l$      \\
	
	$uj$   &   $-\frac{2}{9} (4 g_A^2 + 2 g_1 g_A + g_1^2) ( 2 - 3 q_j) $        
					              &     $(\frac{1}{27} - \frac{1}{18} q_j)g_{\Delta N}^2$     &    $\frac{4}{3} - 2 q_j$      \\
	
	$ul$   &   $-\frac{2}{9} (4 g_A^2 + 2 g_1 g_A + g_1^2) ( 2 - 3 q_l)$       
						      &     $(\frac{1}{27} - \frac{1}{18} q_l)g_{\Delta N}^2$       &    $\frac{4}{3} - 2 q_l$      \\
	
	$dj$   &   $\frac{1}{3} g_1^2 (  1+ 3 q_j)$  &     $( -\frac{1}{27} - \frac{1}{9} q_j)g_{\Delta N}^2$     &    $-\frac{1}{3} - q_j$      \\
	
	$dl$   &   $\frac{1}{3} g_1^2 ( 1 + 3 q_l)$  &     $( -\frac{1}{27} - \frac{1}{9} q_l)g_{\Delta N}^2$     &    $-\frac{1}{3} - q_l$
\end{tabular}
\end{ruledtabular}
\end{table*}

\begin{table*}
\caption{\label{T:SU2n}The coefficients $\b_X$, $\b_X'$, and $A_X$
         in $SU(2)$ flavor \PQCPT\ for the neutron.}
\begin{ruledtabular}
\begin{tabular}{c | c  c  c }
	$X$     &   $\beta_X$   &   $\beta'_X$   &  $A_X$  \\
	\hline
	
	$uu$   &   $\frac{1}{3} g_1^2 (2 - 3 q_j)$  
						      &     $(-\frac{2}{27} + \frac{1}{9} q_j)g_{\Delta N}^2$     &    $-\frac{2}{3} +  q_j$ \\	
	
	$ud$   &   $\frac{4}{9} g_A^2 ( 7 - 6 q_j) - \frac{4}{9} g_1 g_A ( 1 + 3 q_j) + \frac{1}{9} g_1^2( 4 - 6 q_j - 9  q_l)$           
						      &     $(\frac{1}{6} + \frac{1}{18}q_j + \frac{1}{9} q_l)g_{\Delta N}^2$  &    $-1 + 2 q_j + q_l$      \\
	
	$dd$   &   $-\frac{2}{9} (4 g_A^2 + 2 g_1 g_A + g_1^2) ( 1 + 3 q_l)$   &     $( \frac{1}{54} + \frac{1}{18} q_l)g_{\Delta N}^2$     &    $\frac{2}{3} + 2 q_l$      \\
	
	$uj$   &   $-g_1^2 ( \frac{2}{3} -  q_j) $        
					              &     $(\frac{2}{27} - \frac{1}{9} q_j)g_{\Delta N}^2$     &    $\frac{2}{3} -  q_j$      \\
	
	$ul$   &   $-g_1^2 ( \frac{2}{3} -  q_l)$       
						      &     $(\frac{2}{27} - \frac{1}{9} q_l)g_{\Delta N}^2$       &    $\frac{2}{3} -  q_l$      \\
	
	$dj$   &   $\frac{2}{9}(4 g_A^2 + 2 g_1 g_A + g_1^2) ( 1 + 3 q_j)$  &     $( -\frac{1}{54} - \frac{1}{18} q_j)g_{\Delta N}^2$     &    $-\frac{2}{3} - 2 q_j$     \\
	
	$dl$   &   $\frac{2}{9}(4 g_A^2 + 2 g_1 g_A + g_1^2) ( 1 + 3 q_l)$  &     $( -\frac{1}{54} - \frac{1}{18} q_l)g_{\Delta N}^2$     &    $-\frac{2}{3} - 2 q_l$
\end{tabular}
\end{ruledtabular}
\end{table*}


\end{document}